\documentclass[floats,prd,aps,showpacs,amssymb,nofootinbib]{revtex4}
\usepackage{amssymb}
\usepackage{graphics}
\usepackage{amsmath}
\usepackage{amsfonts}
\usepackage{bm}
\usepackage[]{latexsym}
\usepackage{graphicx}
\usepackage{epsf}
\usepackage[]{latexsym}
\usepackage{bm}
\newcommand{\be}{\begin{equation}}\newcommand{\ee}{\end{equation}}
\newcommand{\bea}{\begin{eqnarray}}\newcommand{\eea}{\end{eqnarray}}
\newcommand{\brr}{\begin{array}}\newcommand{\err}{\end{array}}
\newcommand{\bit}{\begin{itemize}}\newcommand{\eit}{\end{itemize}}
\newcommand{\ben}{\begin{enumerate}}\newcommand{\een}{\end{enumerate}}

\newcommand{\ba}{\begin{array}}
\newcommand{\ea}{\end{array}}

\def\lf{\left}

\def\rar{\rightarrow}
\def\ri{\right}
\def\al{\alpha}
\def\De{\Delta}

\def\1{{_{1}}}\def\2{{_{2}}}

\def\noHe0{:\;\!\!\;\!\!:H_e(0):\;\!\!\;\!\!:}
\def\noHm0{:\;\!\!\;\!\!:H_\mu(0):\;\!\!\;\!\!:}

\def\lf{\left}

\def\rar{\rightarrow}
\def\ri{\right}

\def\al{\alpha}
\def\De{\Delta}

\def\1{{_{1}}}\def\2{{_{2}}}

\begin{document}
\title{Heuristic derivation of the Casimir effect from Generalized Uncertainty Principle}

\author{M.~Blasone$^\dagger$\footnote{blasone@sa.infn.it},  G.Lambiase$^\dagger$\footnote{lambiase@sa.infn.it}, G.G.Luciano$^\dagger$\footnote{gluciano@sa.infn.it}, L.Petruzziello$^\dagger$\footnote{lpetruzziello@na.infn.it} and F.Scardigli$^{\ddagger *}$\footnote{fabio@phys.ntu.edu.tw}
}

\affiliation{\mbox{$^\dagger$ INFN and  Universit\`{a} di Salerno,Via Giovanni Paolo II, 132 -- 84084 Fisciano (SA), Italy}\\
\mbox{$^\ddagger$ Dipartimento di Matematica, Politecnico di Milano, Piazza Leonardo da Vinci 32, 20133 Milano, Italy}\\
\mbox{$^{*}$ Institute-Lorentz for Theoretical Physics, Leiden University, P.O.~Box 9506, Leiden, The Netherlands}}

\begin{abstract}
After a short introduction to the generalized uncertainty principle (GUP), we discuss heuristic derivations of the Casimir effect, first from the usual Heisenberg uncertainty principle (HUP), and then from GUP. Results are compared with those obtained from more standard calculations in Quantum Field Theory (QFT).  
\end{abstract}


 \vskip -1.0 truecm
\maketitle
\section{Introduction}
\vspace{2mm}

In his seminal paper of 1927, Heisenberg~\cite{Heisenberg1} introduced the Uncertainty Principle (HUP) in Quantum Theory by discussing measurement processes in which the gravitational interaction between particles was completely neglected. The argument (for example, find the position of an electron by means of photons) is well known under the name of \textit{Heisenberg microscope argument}~\cite{Heisenberg2}. Of course, this procedure was fully justified by the huge weakness of gravity, when compared with the other fundamental interactions. Then things evolved and, in a few years, the HUP became a theorem in the framework of a fully developed Quantum Mechanics~\cite{Robertson}.     

However, if we want to address fundamental questions in Nature, it seems clear that also gravity should be considered when we discuss basic principles and elementary measurement processes. 
This, in fact, is what historically happened, from the very early attempts in generalizing HUP~\cite{GUPearly}, to the more recent proposals like those of string theory, loop quantum gravity, deformed special relativity, non-commutative geometry, and studies of black hole 
physics~\cite{VenezGrossMende,MM,KMM,FS,Adler2,SC-CQG,CGS,SC2013}.

A revised version of the classical Heisenberg argument has been presented in~\cite{FS} and can be described as follows. The size $\delta x$ of the smallest detail of an object,
theoretically detectable with a beam of photons of energy $E$, is roughly given by (if we assume
the dispersion relation $E=pc$)
\be
\delta x
\simeq
\frac{\hbar c}{2\, E}
\ ,
\label{HS}
\ee
so that increasingly large energies are required to explore decreasingly small details.
As remarked above, in its original formulation, Hei\-senberg's gedanken experiment ignores gravity.
Nevertheless, gedanken experiments taking
into account the possible formation, in high energy scatterings, of micro black holes with a
gravitational radius $R_S=R_S(E)$ roughly proportional to the (centre-of-mass) scattering energy $E$ 
(see Ref.~\cite{FS}), suggest that the usual uncertainty relation should be modified as
\be
\delta x
\ \simeq \
\frac{\hbar c}{2\, E}
\ + \
\beta\, R_S(E)
\ ,
\ee
where $\beta$ is a dimensionless parameter.
Recalling that $R_S\simeq 2\,G_N\,E/c^4 = 2\, \ell_p^2\, E/\hbar c$, we can write
\be
\delta x
\ \simeq \
\frac{\hbar c}{2\, E}
\ + \
2\beta\,\ell_p^2\,\frac{E}{\hbar c} \ ,
\label{He}
\ee
where the Planck length is defined as $\ell_p^2=G\hbar/c^3$, the Planck energy as 
$E_p\ell_p=\hbar c/2$, the Planck mass $m_p=E_p/c^2$.
This kind of modification of the uncertainty principle was also proposed in Ref.~\cite{Adler2}.
\par
The dimensionless deforming parameter $\beta$ is not (in principle) fixed by the theory,
although it is generally assumed to be of the order of unity.
This happens, in particular, in some models of string theory (see again for instance
Ref.~\cite{VenezGrossMende}), and has been confirmed by an explicit calculation in Ref.~\cite{SLV}.
However, many studies have appeared in literature, with the aim to set bounds on $\beta$
(see, for instance, Refs.~\cite{brau}).
\par
The relation~(\ref{He}) can be recast in the form of an uncertainty relation, namely a deformation
of the standard HUP, usually referred to as Generalized Uncertainty Principle (GUP),
\be
\Delta x\, \Delta p
\geq
\frac{\hbar}{2}
\left[1
+\beta
\left(\frac{\Delta p}{m_p c}\right)^2
\right]
\ .
\label{gup}
\ee
For mirror-symmetric states (with $\langle \hat{p} \rangle = 0$), the inequality~(\ref{gup}) is equivalent to the commutator
\be
\left[\hat{x},\hat{p}\right]
=
i \hbar \left[
1
+\beta
\left(\frac{\hat{p}}{m_p c} \right)^2 \right]
\ ,
\label{gupcomm}
\ee
since $\Delta x\, \Delta p \geq (1/2)\left|\langle [\hat{x},\hat{p}] \rangle\right|$.
Vice-versa, the commutator~(\ref{gupcomm}) implies the inequality~(\ref{gup}) for any state.
The GUP is widely studied in the context of quantum mechanics~\cite{Pedram},
quantum field theory~\cite{Husain:2012im}, thermal effects in QFT~\cite{FS9506,Scardigli:2018jlm},
and for various deformations of the quantization rules~\cite{Jizba:2009qf}.
   
Casimir effect as well has a long history.
It is a direct manifestation of the non-trivial
nature of quantum vacuum~\cite{BHV98}, which occurs whenever
a quantum field is bounded in a finite region of space; 
such a confinement gives rise to a net attractive force between the confining 
plates, whose intensity has been
successfully measured~\cite{Mohideen:1998iz}.
Since the foundational paper~\cite{Casimir:1948dh},  
the Casimir effect has been largely investigated
in both flat~\cite{Lambiase:1998tf} 
and curved~\cite{Lambiase:2016bjy} background, 
and in particular in quadratic theories of gravity~\cite{Buoninfante:2018bkc}, which
have been studied also in other frameworks~\cite{pscat}. 
Further interesting applications have been addressed in the context
of Lorentz symmetry breaking~\cite{Blasone:2018nfy} and
flavor mixing of fields~\cite{Blasone:2018obn}.

Among the many applications of GUP, some papers have been recently devoted to the study of Casimir effect in the framework of modified uncertainty relations and/or minimal length. In particular, in Refs.~\cite{Nouicer:2005dp,Frassino:2011aa} the corrections to the standard Casimir effect are computed for different GUP scenarios. Such calculations are performed within the framework of QFT, using the formalism of maximally localized states in Hilbert space representation proposed in particular in Refs.~\cite{KMM, DGS}, which implements various types of deformed commutators. 

In the case of GUP involving $ \beta p^2$ term only, as in Eq.~(\ref{gup}), the authors of Refs.~\cite{Nouicer:2005dp,Frassino:2011aa} find corrections to the energy density of the form:
\be\label{casimirgupan}
{\Delta\cal E} = -\frac{\pi^2 \hbar c}{720 d^3}\lf(1 + {\widetilde \beta}
\hspace{0.2mm}\frac{ 2\pi^2\hbar^2 }{3 d^2 }\ri),
\ee
where ${\widetilde \beta}=\beta c^2/E_p^2 $.

On the other hand, recently an interesting derivation of Casimir effect from HUP has been given by Gin\'e~\cite{Gine:2018ncm}  by means of heuristic arguments only.  
Motivated by the utility  of heuristic procedures, which help to develop physical intuition, in this paper we consider similar derivation of the Casimir effect from GUP. We do this for the case of $\beta p^2$ model and in one (space) dimension. 

The work is organized as follows: in Section~\ref{Gine} we briefly review the treatment 
of Ref.~\cite{Gine:2018ncm}. The heuristic derivation of Casimir effect from GUP is given in Section~\ref{GUP}. Conclusions are given in Section~\ref{Conclusions}.

\section{Casimir effect and Uncertainty Principle\label{Gine}}
\vspace{2mm}

In Ref.~\cite{Gine:2018ncm}, an heuristic argument is proposed for deriving the Casimir effect from the HUP. Such a derivation motivated us to look for similar arguments in the case of GUP. 

Here, we briefly review the discussion of Ref.~\cite{Gine:2018ncm}. The Casimir effect is there derived from the idea that the contribution to the energy density at a given point of a plate is modified by the presence of the other plate. 

The author considers  virtual photons produced by vacuum fluctuations arriving at a point $P_0$ of the plate. In principle, the uncertainty $\De x$ about the distance between the photon production point and $P_0$ can have any value within a sphere of (large) radius $R$ enclosing the plates (eventually, such a radius will be sent to infinity). 

The energy density will originate from the contributions of photons from a volume $A \De x$, where 
$A$ is the area of the plate of side $L$.

Now, if there is only one plate, the relevant volume will be the one of entire sphere $V_T=4/3 \pi R^3$ (we take here $R\gg L$). 
The situation is different when another identical parallel plate  is present at distance $d$ from the first one: indeed, in such a case the photons originating from behind the second plate cannot reach the point $P_0$ and, as a result, a reduced  volume $V_T- V_c$ will be associated to the energy density, where $V_c$ is the volume shielded by the second plate (see Ref.~\cite{Gine:2018ncm} for details).

In the case of infinite plates, or even better when $L/(2d) \rar \infty$, one obtains
\be
A \De x = V_T - V_c \simeq \frac{2}{3} \pi R^3\,.
\ee
In the above setting, no length scale is present, so $\De x$ goes to infinity as   
$R$ increases. At this point, the author of Ref.~\cite{Gine:2018ncm} 
introduces a  radius $r_e$, representing the effective distance beyond 
which photons have a very small probability to reach the plate. 

Therefore, the above equation becomes
\be
A \De x = V_T - V_c \simeq \frac{2}{3} \pi r_e^3\,.
\ee
The uncertainty principle in the form 
$\De x \De E \simeq \hbar c/2$, with $E=pc $, is then used to obtain
\be \label{DEgine}
\De E = \frac{3 \hbar c A }{4 \pi r_e^3}\,.
\ee

Finally, the above formula is identified with the exact expression for 
the Casimir energy obtained in QFT:
\be
|\De E| = \frac{\pi^2 \hbar c A }{720  d^3}\,,
\ee
which allows to fix $r_e^3=540 d^3/\pi^3$, i.e. $r_e\simeq 2,6\, d$.

\section{Modified Casimir effect from GUP \label{GUP}}
\vspace{2mm}

The procedure described in the previous Section leading to the Casimir effect from the HUP, appears to be interesting although rather obscure in some points. In particular, in order to clarify the origin of the effective radius $r_e$ above introduced, and to derive the Casimir effect in the GUP framework, we consider here the one-dimensional case.

Starting from the expression of the GUP
\be\label{GUP1}
\De E \De x \simeq \frac{\hbar c}{2}\lf[1 +\beta \lf(\frac{\De E}{E_p}\ri)^2\ri],
\ee
we can certainly write $\De x = \al d$ (being $d$ the only physical scale of the system), where $\al$ is a parameter which will be fixed below.

Eq.~(\ref{GUP1}) can then be rewritten as
\be
\De E = \frac{E_p^2\hspace{0.2mm} \al\hspace{0.2mm} d}{\beta \hbar c}\lf(1 - \sqrt{1 -\beta\lf(\frac{ \hbar c}{E_p \al d}\ri)^2 }\ri)\,.
\ee
This can be expanded  in $\beta\lf(\frac{ \hbar c}{E_p \al d}\ri)^2 \ll 1$, thus  obtaining, up to the first order, 
\be\label{degup}
\De E = \frac{\hbar c}{2\al d}\lf[1 + \beta\lf(\frac{ \hbar c}{2E_p\al d}\ri)^2 \ri].
\ee

Now we fix the parameter $\al$ by taking $\beta\to 0$ in the above equation and comparing the result with the one-dimensional standard QFT expression for the Casimir energy~\cite{Gine:2018ncm}:
\be\label{Casimir1d}
\De E = -\frac{\pi\hbar c  }{12 d}\,,
\ee
from which we get $\al = -6/\pi$. 
Inserting this back into Eq.~(\ref{degup}), we finally obtain
\be\label{casimirgup}
\De E = -\frac{\pi \hbar c}{12 d}\lf[1 + \beta\lf(\frac{ \pi\hbar c}{12 E_p d}\ri)^2 \ri].
\ee
\begin{figure}[h]
 \centering
 \includegraphics[width=11cm]{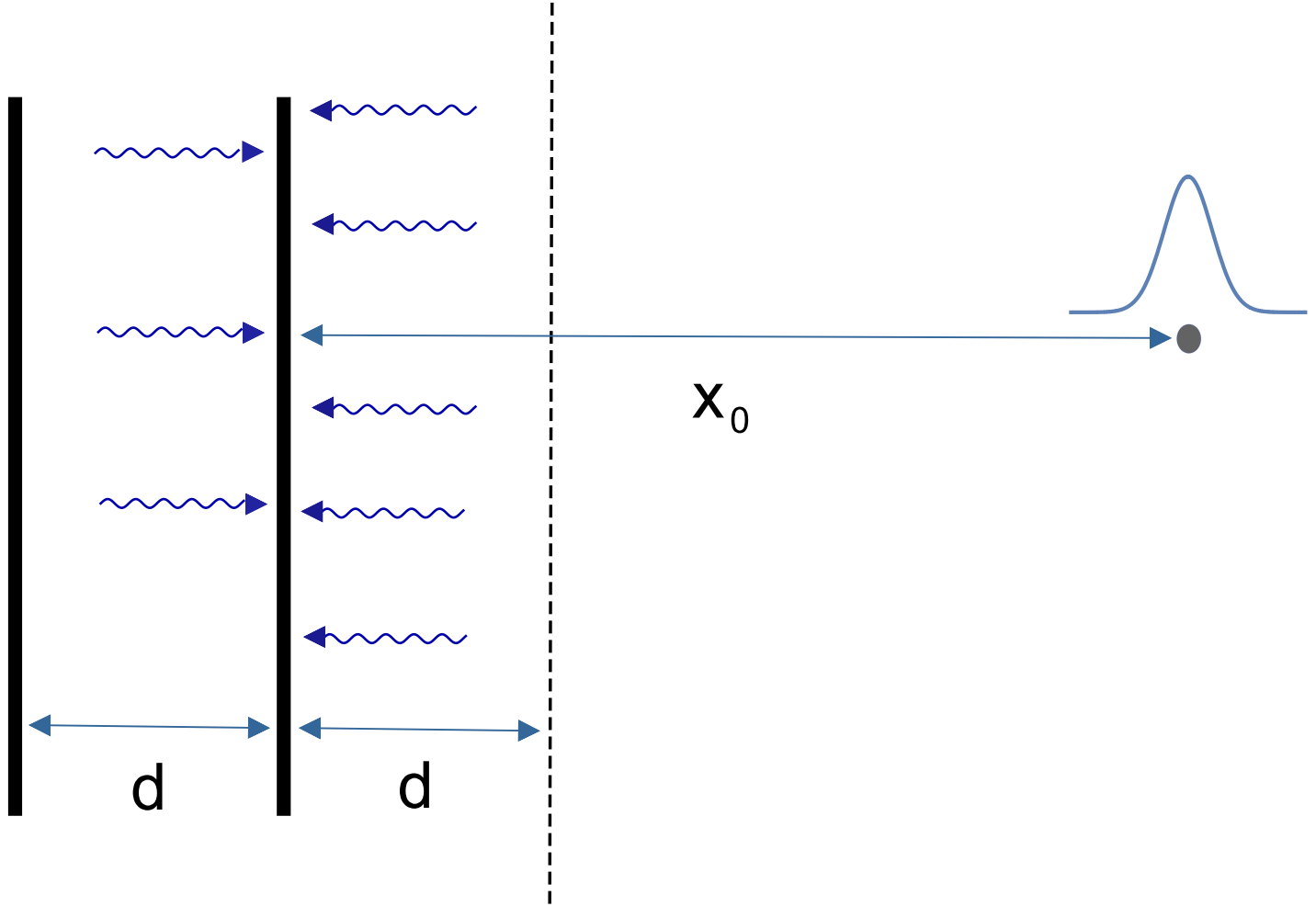}
 \renewcommand{\baselinestretch}{1.0}
 \caption{Setup for the calculation of the Casimir effect: two infinite parallel plates (bold lines), at distance $d$. Photons act on the right plate, coming from right and left regions. See text for detailed explanation.}
 \label{figura}
 \end{figure}
This is to be compared with the expression~(\ref{casimirgupan}) 
of the Casimir energy computed with 
GUP in QFT. In spite of our minimal setting, we
can see that the behavior of GUP correction is approximately recovered\footnote{Strictly
speaking, in order to compare our one-dimensional treatment 
with the three-dimensional one of Ref.~{\cite{Nouicer:2005dp,Frassino:2011aa}},
we should divide Eq.~(\ref{casimirgup}) by $d^2$.
This is the most direct procedure that can be adopted; 
in the case of one-dimensional confinement of the photon field, 
indeed, it is impossible to refer to an energy density per unit area since  
the binding plates reduce to mere points. In spite of these
technicalities, however, we stress that the GUP correction to the Casimir energy
derived via a heuristic approach exhibits the same dependence on $d$
as the more rigorous field theoretical expression in Eq.~(\ref{casimirgupan}).
}.

\section{Conclusions and perspectives}
\label{Conclusions}
\vspace{2mm}

Let us now try to better understand the meaning of the effective radius $r_e$ introduced in Section~\ref{Gine}. It is clear from the Heisenberg uncertainty relation that large fluctuations in energy live for very short times and thus very energetic photons of energy $\Delta E$ can travel only very short distances, of order $\hbar c/(\De E)$. Therefore, energetic photons originating far away from a plate will contribute very little to the energy  around the plate, 
Eqs.~(\ref{DEgine}),~(\ref{Casimir1d}) and~(\ref{casimirgup}).

With reference to Fig.~1, let us focus on the right plate. It is evident that photons originating between the plates, i.e. within a distance $d$ from one plate, will not contribute because of the symmetric action due to the photons originating in the strip of width $d$ on the other side of the plate. On the other hand, photons coming from the right region, from a distance greater than $d$, will not be compensated by photons coming from the left region, whose action is screened by the (infinite) left plate. Therefore only photons coming from a distance greater than $d$, from the right region, can be responsible for a resultant non-zero action on the right plate. Of course, this argument applies symmetrically to the left plate, and here we have a qualitative explanation of the origin of Casimir force. 

Now, consider a point at a distance $x_0>d$ from the plate. Photons are created by quantum fluctuations in a (small) region around that point. The region cannot be smaller than a Compton length of the electron, $\lambda_C=\hbar/m_e c$, otherwise the energy amplitude of the fluctuation would be larger than the threshold ($m_e c^2$) for pairs creation of electrons. Photons can reach the plate (and therefore contribute to the Casimir force) only if their energy is in the range $0<E<E_0$, where $E_0=\hbar c/x_0$. Photons of energies $E$ greater than $E_0$ can only travel paths shorter than $x_0$, before recombining ($x=\hbar c/E < \hbar c/E_0 = x_0$), thus they do not arrive to touch the plate.  Photons at $x_0$ originate from fluctuations of energy $E$ with a probability which
can be assumed to behave like a Boltzmann factor $e^{-E/m_e c^2}$.\footnote{Note that
such an assumption is not fundamental for the validity of the result (\ref{casimirgup}). See the discussion below for more details.} Therefore the total linear energy density (energy per unit length) arriving on the plate from a small region around the point $x_0$ will be
\be
\De \varepsilon \left(E_0\right) = \int_0^{E_0} \lf(\frac{E}{m_e c^2}\ri) e^{-E/m_e c^2} \frac{dE}{\lambda_C} \ = \
\int_0^{E_0} \lf(\frac{E}{\hbar c}\ri) e^{-E/m_e c^2} dE\,,
\ee
where, since we deal with the electromagnetic field, we introduced the natural threshold of the electron mass/energy $m_e c^2$. In terms of $x_0$ the above integral reads
\be
\De \varepsilon(x_0) = \int_{x_0}^{+\infty} \frac{\hbar c}{x^3}\,e^{-\hbar/m_e c x} dx \,.
\label{DeE}
\ee
Finally, summing over all the points $x_0$ such that $d<x_0<+\infty$, we get the whole contribution to the energy from all the photons able to arrive to the plate as
\be
\Delta E(d) = \int_d^{+\infty} \De \varepsilon (x_0)\, dx_0 \,.
\label{DEd}
\ee 
Now, since for $x$ large enough, $e^{-\hbar/m_e c x} \simeq 1$, then Eq.~(\ref{DeE}) yields
\be
\De \varepsilon (x_0) \simeq \frac{\hbar c}{2 x_0^2}\,,
\label{DEx}
\ee
and hence the total energy reads
\be
\Delta E(d) \simeq \frac{\hbar c}{2 d}\, .
\ee
We find therefore again for the energy $\Delta E$ the behavior expressed by Eq.~(\ref{Casimir1d}).

Distributions like those given in Eqs.~(\ref{DEd}),~(\ref{DEx}) allow us to easily define an 
"effective" radius, or distance from the plate, below which the vast majority of photons are responsible for the large part of the Casimir force. In fact, we can define $r_e > d$ as the distance up to which a fraction $\gamma$ ($0<\gamma<1$) of the total energy is responsible for the effect. This means
\be
\int_d^{r_e} \frac{\hbar c}{2 x^2}dx = \gamma \Delta E(d)
\ee  
which yields
\be
r_e = \frac{d}{1-\gamma}\,.
\ee
A final important note is in order at this point. The above equations maintain essentially their validity also for different choices of the distribution function
\be
f(E)=e^{-E/m_e c^2}\,,
\ee  
provided $0<f(E)<1$. In particular, the approximation is as better as the function $f(E)$ is closer to $1$. This is important, since a perusal of current literature has revealed that the search for distributions of the energy fluctuations of the quantum vacuum is not at all a closed chapter, but on the contrary, a hot and magmatic research topic (see for example the important papers of Larry Ford~\cite{LF} and references therein).    

\section*{References}

\end{document}